\documentclass[%
 reprint,
 superscriptaddress,
 twocolumn,
 showpacs,
 preprintnumbers,
 amsmath,amssymb,
 aps,
]{revtex4-1}

\usepackage{graphicx}
\usepackage{dcolumn}
\usepackage{bm}
\usepackage{slashed}
\usepackage{braket}
\usepackage{multirow}
\usepackage{upgreek}
\usepackage{subcaption}
\usepackage{caption}
\usepackage{adjustbox,lipsum}
\usepackage{amssymb}
\usepackage{xspace}  
\usepackage{hyperref}

\def \nobreakseq {\nobreak \hskip 0pt \hbox}

\begin{document}

\newcommand{\Mo}{$\rm ^{100}$Mo\xspace}
\newcommand{\Se}{$\rm ^{82}$Se\xspace}
\newcommand{\Nd}{$\rm ^{150}$Nd\xspace}
\newcommand{\Cd}{$\rm ^{116}$Cd\xspace}
\newcommand{\Sn}{$\rm ^{116}$Sn\xspace}
\newcommand{\Zr}{$\rm ^{96}$Zr\xspace}
\newcommand{\Ca}{$\rm ^{48}$Ca\xspace}
\newcommand{\Te}{$\rm ^{130}$Te\xspace}
\newcommand{\Ge}{$\rm ^{76}$Ge\xspace}
\newcommand{\Ru}{$\rm ^{100}$Ru\xspace}
\newcommand{\Xe}{$\rm ^{136}$Xe\xspace}

\newcommand{\bb}{$\rm 0\nu\beta\beta$\xspace}
\newcommand{\bbnn}{$\rm 2\nu\beta\beta$\xspace}
\newcommand{\DB}{$\rm \beta\beta$\xspace}
\newcommand{\kgy}{kg$\cdot$y\xspace}
\newcommand{\mbb}{$\rm \langle m_{\beta\beta}\rangle$\xspace}
\newcommand{\mee}{$\rm |m_{ee}|$\xspace}
\newcommand{\Qbb}{$\rm Q_{\beta\beta}$\xspace}

\title{Search for Periodic Modulations of the Rate of Double-Beta Decay of \Mo in the NEMO-3 Detector}
 
\setcounter{footnote}{2}

\author{R.~Arnold}
\affiliation{IPHC, ULP, CNRS/IN2P3\nobreakseq{,} F-67037 Strasbourg, France}
\author{C.~Augier} 
\affiliation{LAL, Universit\'{e} Paris-Sud\nobreakseq{,} CNRS/IN2P3\nobreakseq{,} Universit\'{e} Paris-Saclay\nobreakseq{,} F-91405 Orsay\nobreakseq{,} France}
\author{A.S.~Barabash}
\affiliation{NRC ``Kurchatov Institute", ITEP, 117218 Moscow, Russia}
\author{A.~Basharina-Freshville} 
\affiliation{UCL, London WC1E 6BT\nobreakseq{,} United Kingdom}
\author{S.~Blondel} 
\affiliation{LAL, Universit\'{e} Paris-Sud\nobreakseq{,} CNRS/IN2P3\nobreakseq{,} Universit\'{e} Paris-Saclay\nobreakseq{,} F-91405 Orsay\nobreakseq{,} France}
\author{S.~Blot}
\affiliation{University of Manchester\nobreakseq{,} Manchester M13 9PL\nobreakseq{,}~United Kingdom}
\author{M.~Bongrand} 
\affiliation{LAL, Universit\'{e} Paris-Sud\nobreakseq{,} CNRS/IN2P3\nobreakseq{,} Universit\'{e} Paris-Saclay\nobreakseq{,} F-91405 Orsay\nobreakseq{,} France}
\author{D.~Boursette}
\affiliation{LAL, Universit\'{e} Paris-Sud\nobreakseq{,} CNRS/IN2P3\nobreakseq{,} Universit\'{e} Paris-Saclay\nobreakseq{,} F-91405 Orsay\nobreakseq{,} France}
\author{R.~Breier}
\affiliation{FMFI,~Comenius~University\nobreakseq{,}~SK-842~48~Bratislava\nobreakseq{,}~Slovakia}
\author{V.~Brudanin} 
\affiliation{JINR, 141980 Dubna, Russia}
\affiliation{National Research Nuclear University MEPhI, 115409 Moscow, Russia}
% ...
\collaboration{NEMO-3 Collaboration}
\noaffiliation

\date{\today}


\begin{abstract}


Double-beta decays of \Mo from the 6.0195-year exposure of a 6.914\,kg high-purity sample were recorded by the NEMO-3 experiment that searched for neutrinoless double-beta decays. These ultra-rare transitions to \Ru have a half-life of approximately $7\times10^{18}$\,years, and have been used to conduct the first ever search for periodic variations of this decay mode. The Lomb-Scargle periodogram technique, and its error-weighted extension, were employed to look for periodic modulations of the half-life. Monte Carlo modeling was used to study the modulation sensitivity of the data over a broad range of amplitudes and frequencies. Data show no evidence of modulations with amplitude greater than 2.5\% in the frequency range of $0.33225\,{\rm y^{-1}}$ to $365.25\,{\rm y^{-1}}$. 

\end{abstract}



\maketitle
\section{\label{sec:introduction}Introduction}

The invariance of fundamental constants of nature has been scrutinized in a broad range of physics contexts including considerations discussed by Milne~\cite{Milne}, Walker~\cite{Walker}, Dirac~\cite{Dirac-Nature, Dirac:1938mt}, Chandrasekhar~\cite{Chadrasekhar}, and Kothari~\cite{Kothari} during the initial ascent of data-based cosmology of the expanding Universe. Modern cosmology and the observed evolution of the Universe are closely related to properties of interactions of elementary particles, and impose tight bounds on possible changes in constants that include the strength of gauge couplings as a function of time elapsed since the Big Bang. A number of authors studied these constraints and their implications, e.g.,~\cite{Weinberg-1983a, Weinberg:1983xy, Kolb:1985sj, Preskill:1988na, Olive:2002tz, Olive:2011xvo, Uzan:2015uba, Fritzsch:2016ewd, Braconi:2018gxo, Balcerzak:2019smn}. Many theoretical ideas and implied phenomena are discussed in literature and have been reported in reviews~\cite{Uzan:2010pm, Martins:2017yxk} that also contain exhaustive lists of references on this subject.

Closely related to the time invariance of constants is their periodicity, given the ubiquitous presence of cyclic processes in nature at almost all distance and time scales. There are ongoing searches and tests of such phenomena, mostly but not only connected to dark matter and dark energy~\cite{Lewin:1995rx, Freese:2012xd, Mayet:2016zxu}. Some experiments have produced controversial results suggesting yearly modulation of WIMP interactions, e.g.,~\cite{Bernabei:2010mq, Bernabei:2008yi, Bernabei:2018yyw, Aalseth:2010vx, Aalseth:2011wp} as an expected `smoking gun' observable in direct searches of dark matter~\cite{Drukier:1986tm}. However, these claims of  low-mass WIMP dark matter  are  not supported by other experiments~\cite{Aprile:2018dbl, Aprile:2019jmx, Akerib:2016vxi, Agnese:2018gze, Agnes:2018oej}.  Also controversial are results of {\it a posteriori} data analyses of measurements of nuclear decay half-lives which have yielded unexpected periodicities, including annual modulations that authors have linked to the periodicity of the Earth--Sun distance and solar activity~\cite{Jenkins:2008tt, Jenkins:2008vn, Fischbach:2009zz, Javorsek:2010sr, Sturrock:2010, Sturrock:2010bu, Sturrock:2014caa, Sturrock:2012gs, Jenkins:2012jc, Sturrock:2012re, Sturrock:2019dfx, parkhomov}. However, other analyses of the same data do not reveal any significant modulations~\cite{Pomme:2017dfi, Pomme:2019mhq, Dhaygude:2019hka}.

Difficulties with precision testing the time variation of half-lives of long-lived radioisotopes can be attributed, in part, to two important factors: the time duration of measurements, and  the necessity of long-term control of background phenomena, which often exhibit time and/or seasonal periodicities. Examples of such phenomena include the average environmental temperature; radon levels in the ground, buildings, and caverns; the seasonal cosmic-ray flux modulation; the solar wind intensity related to the Earth--Sun distance; the phase of the lunar cycle and tides; solar activity, etc. 

In this Letter, we are reporting results of a search for periodicity of double-beta (\DB) decays of \Mo$\rightarrow$\Ru. Data were collected over a period of approximately eight years by the NEMO-3 experiment~\cite{NEMO-3:2019gwo, Arnold:2015wpy}. The experiment, designed to search for neutrinoless double beta decay (\bb),  had an exquisite capability of background identification and suppression. NEMO-3 collected an unprecedentedly large data set of 2-neutrino double-beta decays (\bbnn). These extremely rare events with two electrons in the final state of the \Mo decay, whose half-life is about $7\times10^{18}$\,years, served as a unique testing ground for the first ever search for periodicity of a second-order weak transition on time scales shorter than or comparable to the measuring period. 

\section{\label{sec:NEMO-3}The NEMO-3 Experiment}

The NEMO-3 detector~\cite{Arnold:2004xq} was designed to detect two electrons in the final state of neutrinoless double-beta decays. Thin foils of the source isotopes were surrounded by a tracking chamber and plastic calorimeter blocks that reconstructed the full kinematics of various decays and interactions within the detector. The source foils were strips about 65\,${\rm mm}$ wide, 2,480\,${\rm mm}$ long, and $40-60$\,${\rm mg/cm{^2}}$ thick and were made of various double-beta decay isotopes (\Mo, \Se, \Te, \Cd, \Nd, \Zr, and \Ca) totaling about 10\,kg. 

These foil strips were arranged vertically to form a cylinder such that the tracking and calorimetric volumes on either side formed a toroidal geometry for the whole detector. For the purposes of this analysis, only events from the \Mo foils were considered. This particular isotope constituted the majority, 6.914\,kg, of the total source mass in NEMO-3. The tracking volume that surrounded either side of the foils was comprised of 6,180 drift wire cells, operating in Geiger mode, within a gas mixture of helium-argon (95\%-1\%), and ethanol with water vapor (4\%). The tracking volume was then further enclosed by the calorimeter walls composed of 1,940 plastic scintillator blocks coupled to photomultiplier tubes. Finally, a large solenoid encircled the detector to produce a 25\,${\rm G}$ magnetic field to help with e$^+$/e$^-$ discrimination. The entire detector was shielded from external backgrounds by a combination of iron, wood, paraffin, and borated water. The detector was placed in the Modane Underground Laboratory in the Fr\'{e}jus tunnel in the Alps which provided 4800 m.w.e. overburden to shield from cosmic rays.

The detector was operated from early 2003 until early 2011 with data taking split into two run periods known as Phase 1 (February 2003 - September 2004) and Phase 2 (October 2004 - January 2011). During Phase 2, an additional  enclosure was installed, surrounding the detector. Filled with radon-filtered air, this enclosure greatly reduced radon permeation into the detector. The result was a significant increase in the purity of the signal channel for Phase 2. Only these lower background runs were considered for the final analysis presented herein. The total span of the runs from this period amounts to $6.0195$\,${\rm y}$.
\section{\label{sec:selection}Event Selection}

The purpose of this analysis was to search for periodic trends in the rate of double-beta decays (with no distinction between \bbnn and theoretical \bb decays) originating in the \Mo{} source foils, which required counting the number of such decays per unit time recorded in the NEMO-3 detector. Data taking was divided into specific run periods with durations ranging from tens of minutes to just over two days. The \Mo activity yielding double-beta events was about $0.1$\,${\rm Bq}$ and, since most runs lasted longer than 20 minutes, most of them accumulated over 100 double-beta events. A discrete time series of the decay rate was constructed by taking the number of selected events in a run divided by the duration of the run to yield a value for the observed rate. The average double-beta event rate for each run was associated with a timestamp, corresponding to the midpoint of the run.

Events that were selected for inclusion into this calculation were chosen based on a wide range of criteria to minimize the contribution from background processes which could mimic topologies of a double-beta decay. The primary characteristics of a double-beta decay event are the identification of two tracks with curvatures consistent with negatively-charged particles, originating from the \Mo foil, and
with associated energy deposits in scintillator blocks; with no coincident alpha particles (short straight tracks); and no gamma particles (unassociated scintillator hits) with energy deposits greater than or equal to $150$\,${\rm keV}$. 

The electron track lengths were required to sum to at least $60$\,${\rm cm}$ and trace back to a common vertex in one of the \Mo source foils (with no more than $4$\,${\rm cm}$ of difference between the two tracks either transverse to, or in the plane of the foils). Vertices were required to fall outside of ``hot spots'' -- regions that were identified as contaminated with radio-impurities. The electrons were required to fire Geiger cells within $50$\,${\rm cm}$ of the vertex on the foil. Electron energy deposits were required to exceed $200$\,${\rm keV}$ in each of two separate scintillator blocks, each of which had no neighboring blocks with energy deposits. The extrapolated electron tracks were required to hit the front face of each block, and each block had to be coupled to a PMT that had not been flagged with potential issues such as excess noise or a lack of calibration data. Electrons that hit the blocks nearest to the foil on the endcaps (scintillator blocks at the top and bottom of the detector)  were also rejected due to a higher chance of incorrectly identifying the track curvature and thus particle charge. 

\begin{figure}[!htbp]
  \includegraphics[width=0.4\textwidth]{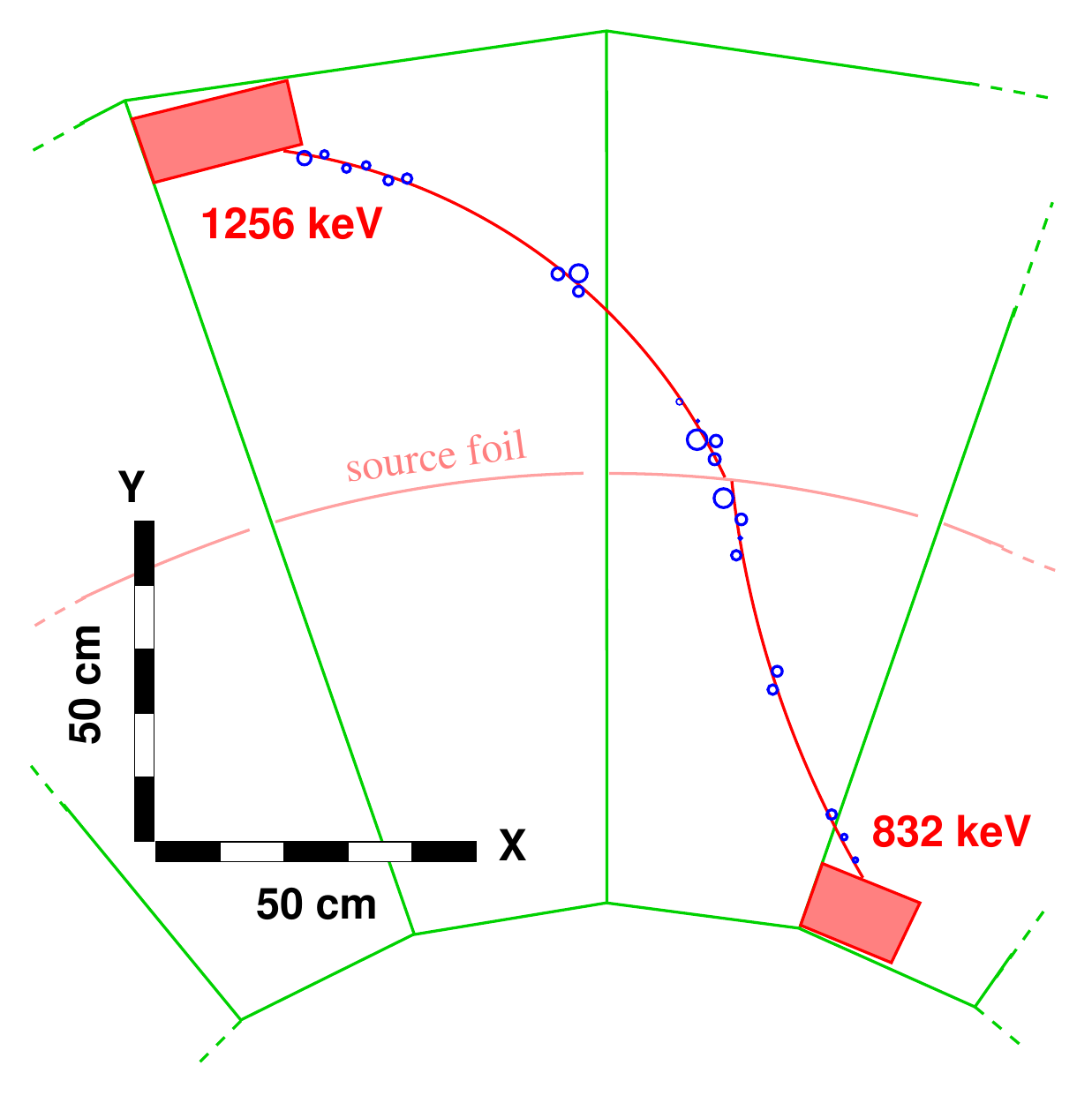}
  \caption[Event Display]{\small{An example event display from NEMO-3 showing hit Geiger cells (blue circles) and calorimeter modules (red boxes). A helix has been fit to the hit cells to show the curvature of the two electron tracks (due to the magnetic field) and to show that they share a common origin on the source foil.}}
  \label{fig:EventDisplay}
\end{figure}

Time-of-flight measurements obtained from calorimeter hit times were also used to calculate the probability of the event originating within the foil, compared to being an external event in which a particle passes through the foil. The event had to satisfy the condition that the probability of originating within the foil was greater than or equal to 4\% while the probability of being an external crossing event was less than or equal to 1\%. Only runs that were deemed to be of good quality were used. An event display for a candidate double-beta decay which passed all such selection criteria is shown in Figure~\ref{fig:EventDisplay}. This event selection resulted in a very high sample purity (a very similar event selection resulted in a signal to background ratio of 76~\cite{Arnold:2013dha}) and no background subtraction was performed in calculating per-run event rates. The possibility of time varying backgrounds was also considered, in particular due to the potential seasonal variation of radon levels. No modulations appeared in the analysis, of either the Phase 1 or Phase 2 data, at frequencies corresponding to such processes.


%
The decay rate based on the raw number of events that passed these selection requirements had to be corrected by the efficiency of the detector during each run period. This correction scaling could vary from one run to the next and was accounted for using precise and comprehensive Monte Carlo simulations of the NEMO-3 experiment. 

The final corrected \Mo double-beta decay rate time series as measured by the NEMO-3 detector is shown in the upper plot of Figure~\ref{fig:Results}. The data yielded an average per-run rate of $0.085$\,${\rm Hz}$ with a standard deviation of 11.8\%. The average run was 8.3 hours in duration (standard deviation of 4.7 hours) and saw approximately 115.9 events in that time (standard deviation of 66.7 events). In total, 449,733 events were collected, with a mean efficiency of 4.6\%, across the 3,869 run periods. A study of the most significant systematic errors showed that their contributions were small compared to statistical fluctuations from one run to the next.

%
%
\section{\label{sec:Analysis}The Search for Periodicities}

A common approach to searching for periodic trends in data involves decomposing the time series of interest into its spectral frequency components. In one such approach, a periodogram (also called a power spectrum) can be constructed by calculating, for any number of desired sample frequencies, a quantity known as the power. Frequencies which produce a large power are those which have a stronger presence in the underlying data set. The basic periodogram technique developed by Lomb~\cite{lomb} was eventually modified to allow for unevenly sampled data to be analyzed and became known as the Lomb-Scargle (LS) periodogram~\cite{scargle}. Further developments eventually allowed the technique to also account for weighted data and an overall offset term, known as the Generalized Lomb-Scargle (GLS) periodogram~\cite{zechmeister}. 

Periodogram analyses, which are commonplace in astronomy and astrophysics, have also seen use in the fields of nuclear and particle physics. Some of the previously mentioned dark matter searches, e.g.,~\cite{Bernabei:2010mq, Bernabei:2008yi, Bernabei:2018yyw} and radioactive decay analyses, e.g.,~\cite{Javorsek:2010sr, Sturrock:2010, Sturrock:2010bu, Sturrock:2014caa} and~\cite{parkhomov} have used LS periodogram analyses similar or identical to the approach used herein. Other examples also include searches for periodic variations in neutrino fluxes across various experiments, e.g.,~\cite{Milsztajn, yoo, sturrock, Sturrock:1999, aharmim, Altmann}, while uses of the GLS periodogram appear in other recent analyses~\cite{Tejas, Dhaygude:2019hka, Nakano}.

For a discrete time series $X(t_j)$ consisting of $N$ entries, the basic Lomb-Scargle power, $P_{\text{LS}}(\omega)$ can be calculated at a given sample frequency, $f$ (where $\omega=2 \pi f$), by 
\begin{equation} \label{eq:PLS}
  \begin{split} 
    P_{\text{LS}}(\omega) = \frac{1}{2 \sigma^2}&\left\{\frac{\left(\displaystyle\sum_{j=1}^{N} [X(t_j)-\bar{X}]\cos[\omega(t_j - \tau)]\right)^2}{\displaystyle\sum_{j=1}^{N} \cos^2[\omega(t_j - \tau)]}\right. \\
    &+ \left.\frac{\left(\displaystyle\sum_{j=1}^{N} [X(t_j)-\bar{X}]\sin[\omega(t_j - \tau)]\right)^2}{\displaystyle\sum_{j=1}^{N} \sin^2[\omega(t_j - \tau)]}\right\}, 
  \end{split} 
\end{equation} 
where $\bar{X}$ is the mean of the data points, $\sigma$ is their standard deviation, and $\tau$ is defined by the relation $\tan(2\omega\tau)=\sum_{j=1}^{N}\sin(2\omega t_j)/\sum_{j=1}^{N}\cos(2\omega t_j)$. The periodogram is constructed by calculating this power over a range of frequencies of interest. While powerful in its handling of unevenly spaced data, the LS technique doesn't weight each data point by its uncertainty. 

A well known property of the LS technique is its equivalence to least-squares fitting of sine waves~\cite{scargle}. By taking into account an offset term and weights, the GLS technique extends this equivalence to a full ${\chi}^2$ fitting approach. The end result is a new expression for the Generalized Lomb-Scargle power, $P_{\text{GLS}}(\omega)$, given by
\begin{equation} 
  \begin{split} \label{eq:PGLS}
    P_{\text{GLS}}(\omega) = \frac{1}{XX \cdot D}&\left[ SS \cdot (XC)^2 + CC \cdot (XS)^2 \right. \\
    &\left.- 2CS \cdot XC \cdot XS \right] 
  \end{split}
\end{equation}
where $D = CC \cdot SS - (CS)^2$ and the following abbreviations are used (with summations running over the same indices as in Eq.~(\ref{eq:PLS})): \newline
$XX = \sum w_j [X(t_j)-\bar{X}]^2,\\ XC = \sum w_j [X(t_j)-\bar{X}]\cos(\omega t_j),\\ XS = \sum w_j [X(t_j)-\bar{X}]\sin(\omega t_j),\\ CC = \sum w_j \cos^2(\omega t_j) - [\sum w_j \cos(\omega t_j)]^2,\\ SS = \sum w_j \sin^2(\omega t_j) - [\sum w_j \sin(\omega t_j)]^2,\\ CS = \sum w_j \cos(\omega t_j)\sin(\omega t_j) - [\sum w_j \cos(\omega t_j) \times \sum w_j \sin(\omega t_j)]$. \newline
Here the $w_j$ are the weights for each $X(t_j)$, given by $w_j = \frac{1}{W}\frac{1}{\sigma_j}$ for $W=\sum\frac{1}{\sigma_j}$ where the $\sigma_j$ are the errors, and so the mean is now $\bar{X} = \sum w_j X(t_j)$. 
%
%

Another known feature of the LS periodogram is that if the time series $X(t_j)$ are made up of Gaussian random values with no underlying modulation then the resultant periodogram powers should be exponentially distributed with unit mean~\cite{scargle}. This allows one to estimate the false-alarm probability (F.A.P.), which gives the probability of finding a power larger than $P$, via the expression
\begin{equation} \label{eq:FAP}
  \text{F.A.P.}(P) = 1 - (1- e^{-P})^M,
\end{equation}
where $M$ is the number of frequencies sampled. This lets one calculate a percentage confidence level (C.L.) value for a given power as
\begin{equation} \label{eq:CL}
  \text{C.L.}(P) = (1 - e^{-P})^M \times 100\%.
\end{equation}
This approach to estimating the significance of periodogram peaks is very straightforward but depends critically on the initial assumption of the Gaussian distribution of the data points. Furthermore, the GLS periodogram expression given in Eq.~(\ref{eq:PGLS}) requires a normalization factor, for which the authors of~\cite{zechmeister} offer multiple approaches, in order for Eq.~(\ref{eq:CL}) to apply. Here, the normalization scheme proposed by Baluev~\cite{baluev} was used so that LS and GLS powers could be shown on the same scale. However, due to these added complications, the most reliable way to assess the significance of periodogram peaks is via Monte Carlo methods which are described in section~\ref{sec:MC}. The simple but rough significance approximation obtained from Eq.~(\ref{eq:CL}) was thus employed for sensitivity studies which were optimized to reduce computational loads.

Both the LS and GLS techniques were employed in this analysis. The two were used as both a cross-check to one another as well as for consistency when comparing to other analyses which predominantly relied on the more commonly-used LS technique. To prevent analysis bias, a blinded approach was taken by first applying both techniques to time-shuffled Phase 1 data and then to unshuffled Phase 1 data before finally applying them to the Phase 2 time series. Here, the term shuffling is used to describe a randomized re-ordering of data points so as to destroy any potential underlying temporal trends in the data. 

\begin{figure*}[!htbp]
  \includegraphics[width=0.95\textwidth]{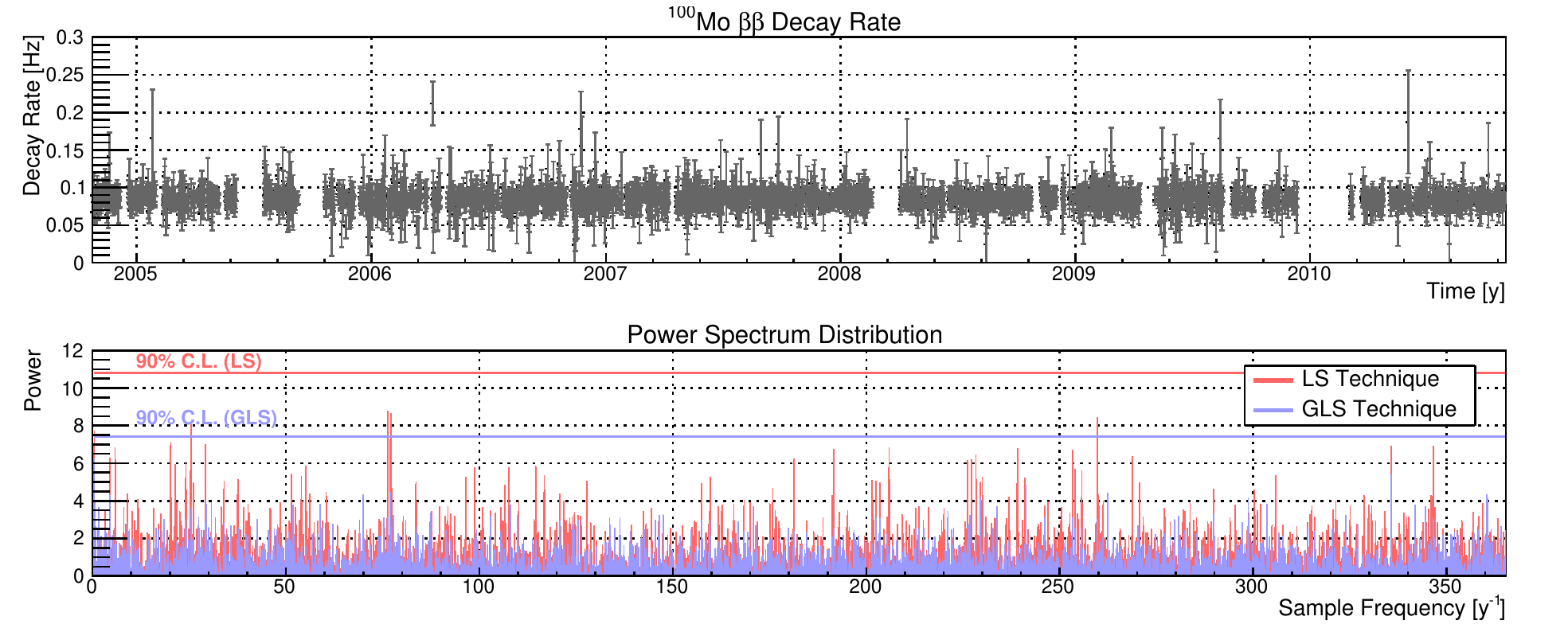}
  \caption[Power Spectrum Analysis of \Mo{} double-beta Decay Rate - Full Frequency Range]{\small{Results of applying both of the periodogram techniques to the double-beta decay rate time series of Phase 2 \Mo{} data. The upper plot shows the efficiency-corrected time series that was analyzed. The lower plot shows the periodograms obtained from applying the LS and GLS techniques. The horizontal lines denote the threshold powers, in each technique, for reaching a 90\% confidence level based on the ``shuffle test''  method described in Section~\ref{sec:MC}.}}
  \label{fig:Results}
\end{figure*}

The final time series contained 3,869 runs and an oversampling factor of two was used. This meant that 7,738 frequencies, twice the number of data points, were sampled. These frequencies were evenly distributed in the range $[0.33225,365.25]\,\text{y$^{-1}$}$ which was chosen based on the maximum and minimum modulation periods that could be detectable within the given duration of data taking. The minimum period was limited by the average run spacing ($\Delta T_{avg}$) such that $f_{\text{max}} = 1/(2 \times \Delta T_{avg})$. A conservative value of 12 hours was used for the average run spacing which translates to a minimum period of one day or $f_{\text{max}} = 365.25\,\text{y$^{-1}$}$. Similarly, the maximum period was determined by the total span of the data. In this case, the minimum sample frequency was chosen such that at least two full periods of a modulation would be contained in the data which implies that $f_{\text{min}} = 2/(\text{span}) = 0.33225\,\text{y$^{-1}$}$. The resultant LS and GLS periodogram are shown in the lower plot of Figure~\ref{fig:Results}. 

Upon constructing the LS and GLS periodograms for the data, the largest LS power of 8.78 was found at the frequency of $76.26\,\text{y$^{-1}$}$, corresponding to a periodicity of approximately 4.8 days, while the largest GLS power of 6.16 was found at the frequency of $0.47\,\text{y$^{-1}$}$, corresponding to a periodicity of approximately 777 days or 2.1 years. The fact that the locations of these peaks disagreed between the two techniques, and that neither had any correlation with periodicities found in any of the previously mentioned references, indicates they are very likely truly random fluctuations. As further evidence, the GLS power found at the same frequency as the largest LS power is smaller than any of the top three largest GLS powers, and vice versa, wich emphasizes how uncorrelated the maximal peaks are between the two techniques. The sizes of the largest periodogram peaks are analyzed in more depth in the next section to further corroborate this assertion. 
%
\section{\label{sec:MC}Significance and Sensitivity Studies}

A common approach to determining the significance of periodogram peaks is to test how often a specific power is exceeded in a similar time series comprised of randomized pseudo-data with no modulation. The same shuffling procedure, described in the previous section to blind the Phase 1 data, was applied to the Phase 2 data 10,000 times, to create a collection of randomized, null-hypothesis time series. This preserved the structure of the original data in terms of the actual values and their associated errors as well as their temporal spacing. The LS and GLS techniques were then applied to each of these time series and the largest powers were recorded for each resultant periodogram. The largest powers from the true data could then be compared against these maximal powers from the pseudo-data to estimate their significance. 

\begin{figure}[!htbp]
  \centering 
    \begin{subfigure}[b]{\linewidth}
      \includegraphics[width=0.85\linewidth]{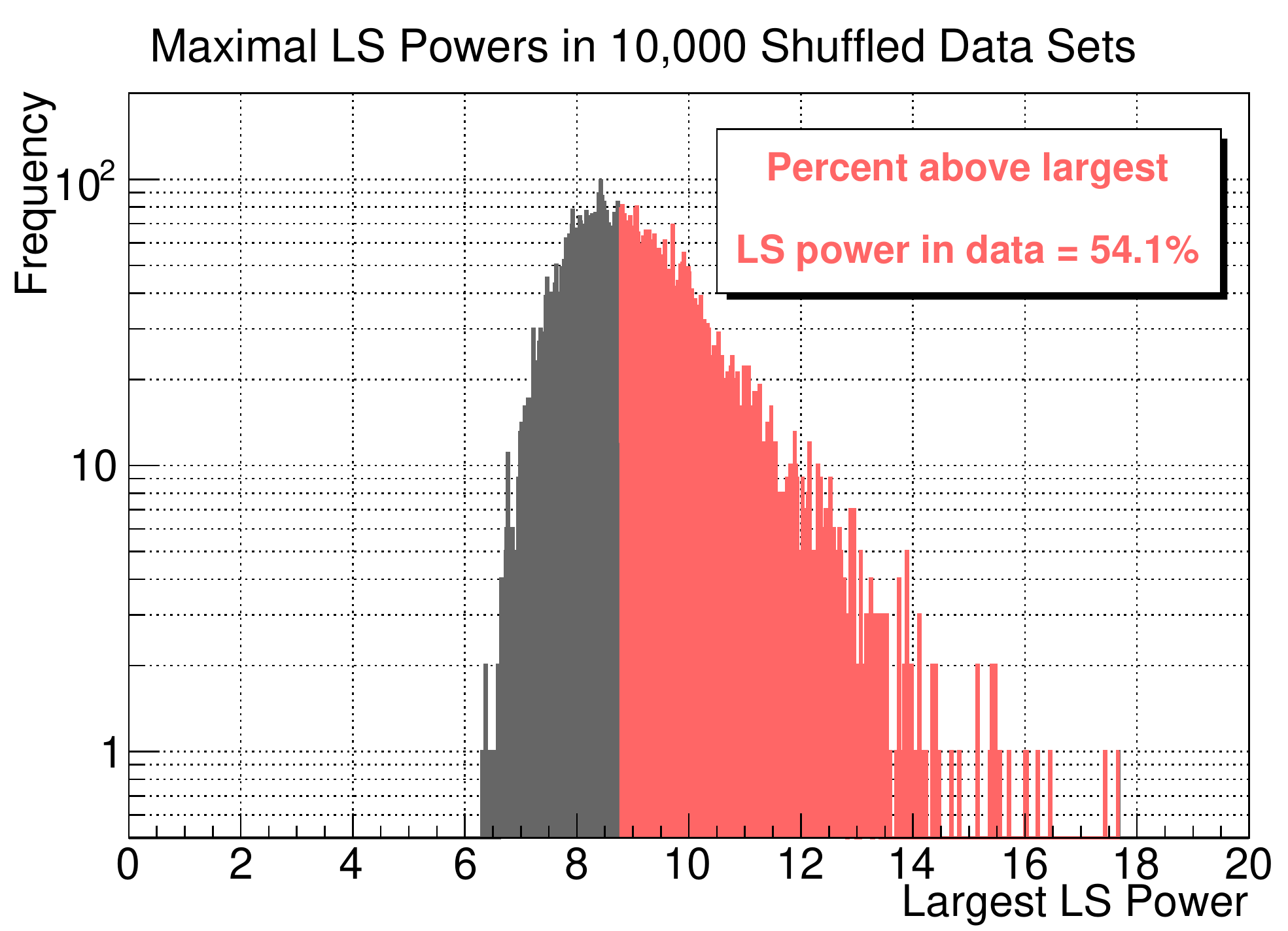}
      \label{fig:NullHypLS}
    \end{subfigure}
    \hfill
    \begin{subfigure}[b]{\linewidth}
      \includegraphics[width=0.85\linewidth]{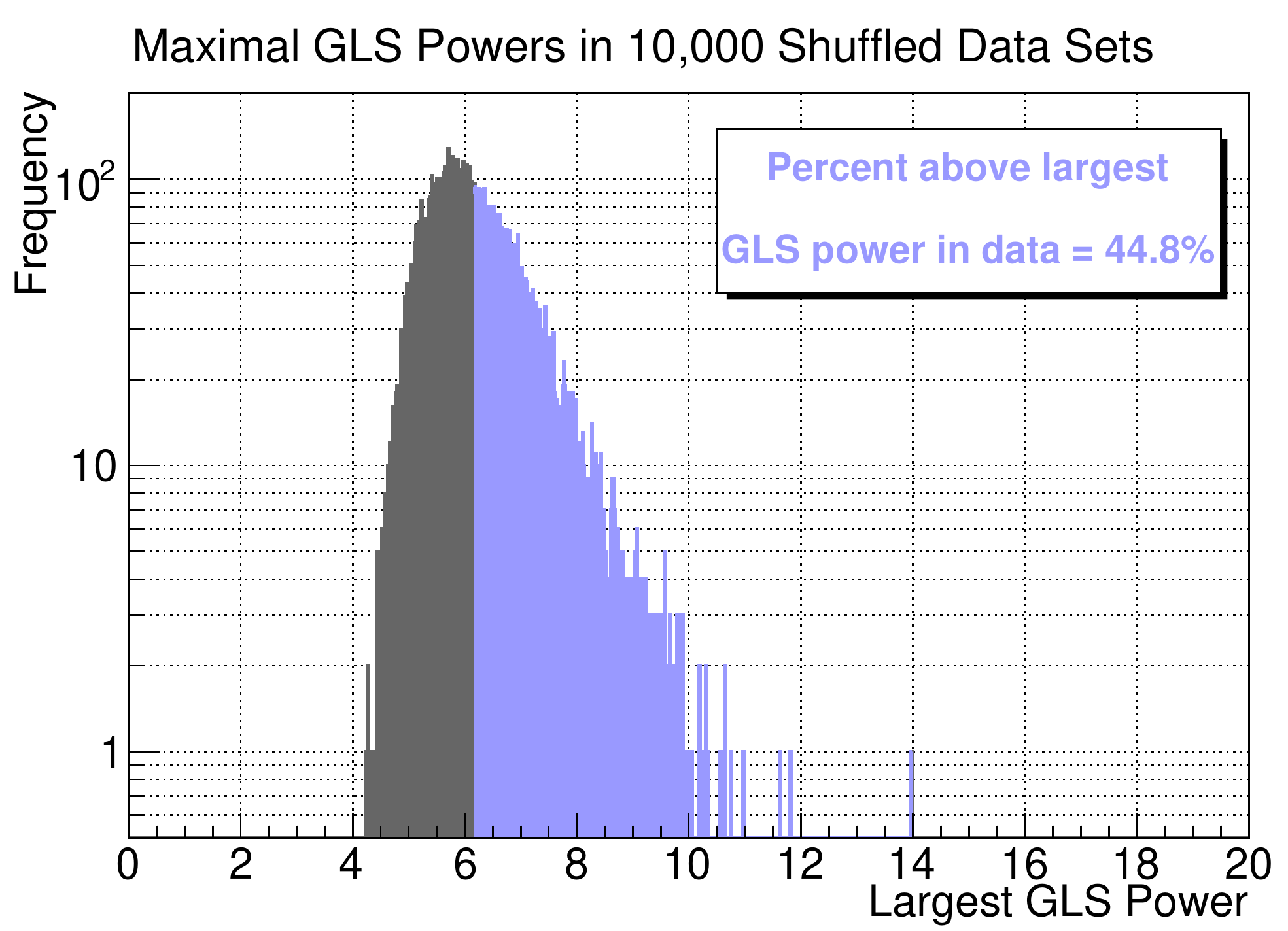}
      \label{fig:NullHypGLS}
    \end{subfigure}
    \caption[Null Hypothesis Testing]{\small{A distribution of the maximal LS (top) and GLS (bottom) powers in 10,000 randomly shuffled null-hypothesis data sets (no applied modulations). The fraction of data sets with maximal power greater than the maximum LS (GLS) power observed in the data constitute 54.1\% (44.8\%). This estimates the significance of the observed maximum LS (GLS) power in the data to be at the 45.9\% (55.2\%) confidence level.}}
  \label{fig:NullHyp}
\end{figure}

This so called ``shuffle test'' method has been used in a similar decay rate analysis by Sturrock~\cite{Sturrock:2010} and has also been attributed to an analysis searching for modulations in the Homestake solar neutrino experiment~\cite{bahcall}. The results of applying the shuffle test for the LS and GLS techniques are shown in Figure~\ref{fig:NullHyp}. The division in each distribution shows what percentage of maximal powers lie above the largest power found in the data. For the LS (GLS) technique where the largest power was 8.78 (6.16), a larger power was found in the shuffled pseudo-data sets 54.1\% (44.8\%) of the time. 
%

Further studies with these Monte Carlo data sets were also undertaken to estimate the sensitivity of the data to detecting different modulations. To do this, a modulation to the pseudo-data sets was applied in the form of 
\begin{equation}
  R(t_i) = N[1 + A \times \sin (2 \pi f t_i + \phi)],
\end{equation}
where $N$ is a normalization constant to match the mean of the un-modulated data; $A$ represents a fractional or relative amplitude (which will be denoted as a percentage relative to the mean rate) for the applied modulation; $f$ is the frequency of the applied modulation; and $\phi$ is the modulation phase. 

Although the final analysis only sampled frequencies corresponding to modulations in which at least two full periods are present in the data, the sensitivity studies sampled down to even smaller frequencies. This helped determine an apt lower bound and allowed for the study of the behavior of the two techniques in the low frequency domain. Due to the computationally intensive nature of these studies, the sample frequency range was broken down into three regimes (low, mid, and high) that included different spacing between frequencies to limit the total number of trials that were needed. A summary of the various amplitudes and frequencies that were used for the injected modulations are shown in Table~\ref{tab:TestFreqs}. Early studies showed that the modulation phase only affected the sensitivity contours in the low frequency regime (which was irrelevant to the final analysis parameters). Thus the phase was set to zero in generating the results shown below (more will be discussed about the effects of the phase value further on).
\begin{table}[!htbp]
  \begin{center}
    \begin{tabular}{  c | c | c  }

      & \textbf{Amplitude~($A$)~[\%]} & \textbf{Step Size ($\Delta A$)~[\%]}  \\
      \hline
      & 0.5 - 4 & 0.1  \\
      \hline
      \textbf{Range} & \textbf{Freq.~($f$)~[y$^{-1}$]} & \textbf{Step Size ($\Delta f$)~[y$^{-1}$]} \\
      \hline
      Low & 0.03 - 0.1 & 0.005 \\
      Mid & 0.15 - 2 & 0.05 \\
      High & 15 - 360 & 15 \\

    \end{tabular}
    \caption[Summary of the Amplitude and Frequency Parameters Used for Sensitivity Studies]{\small{Summary of the amplitude and frequency ranges and spacings used for the injected modulation signals to test for detection sensitivity.}}
    \label{tab:TestFreqs}
  \end{center}
\end{table}

\begin{figure*}[!htbp]
  \centering 
    \begin{subfigure}[b]{\linewidth}
      \includegraphics[width=\linewidth]{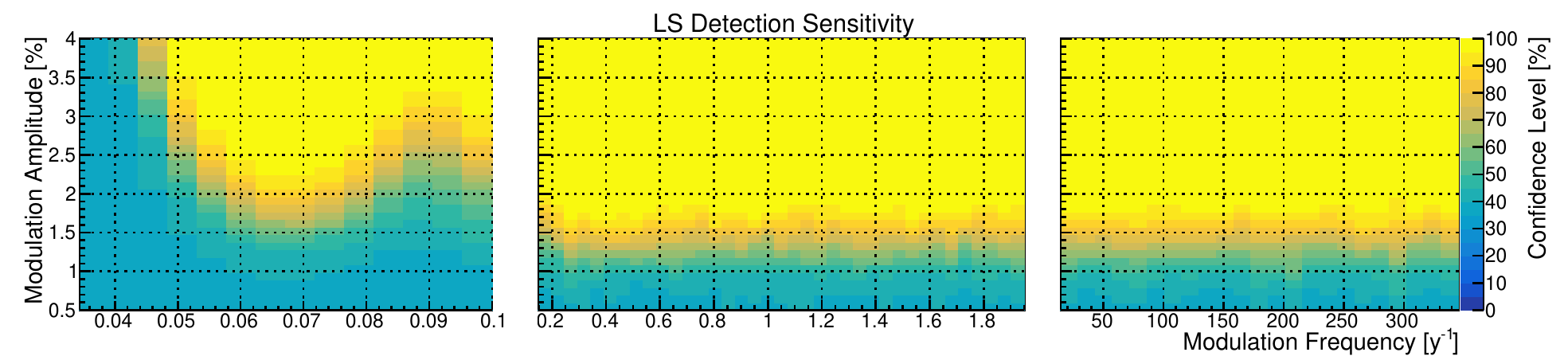}
      \label{fig:SensitivityLS}
    \end{subfigure}
    \hfill
    \begin{subfigure}[b]{\linewidth}
      \includegraphics[width=\linewidth]{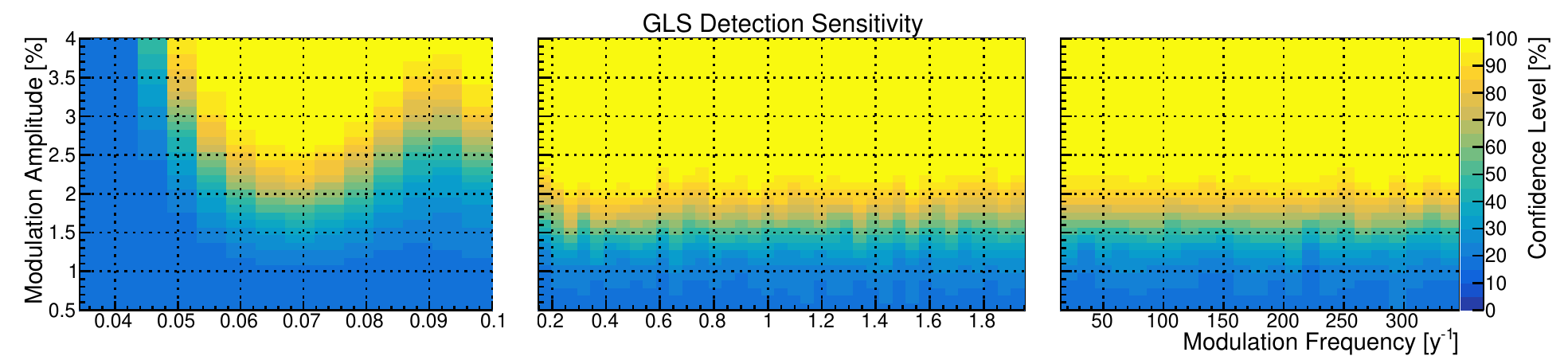}
      \label{fig:SensitivityGLS}
    \end{subfigure}
    \caption[Sensitivity Contours]{\small{An illustration of how the LS and GLS detection sensitivity changes as the input modulation parameters $A$ and $f$ are varied (for $\phi = 0$ and three different frequency regimes - low, mid, and high). The z-axis, color, at each point denotes the estimated significance (in terms of C.L. value) of the largest peak, averaged over 100 periodograms.}}
  \label{fig:Sensitivity}
\end{figure*}

For each point in this amplitude-frequency phase space 100 different Monte Carlo pseudo-data sets were analyzed to average out random variations. At each combination of modulation amplitude and frequency, the LS and GLS periodograms were constructed for the different pseudo-data sets and the average (across all 100 sets) of the largest power was recorded, as well as its estimated C.L. value derived from Eq.~(\ref{eq:CL}). This was used to create a contour plot of C.L. values in the modulation amplitude-frequency space. These are shown, for the three frequency regimes, in Figure~\ref{fig:Sensitivity} for both techniques.
%
%
%

For mid and high ranges of modulation frequencies, the detection sensitivity only depended on the modulation amplitude. In these regimes the threshold for detection at approximately a 95\% C.L. was about 2.0\% (2.5\%) relative amplitude for the LS (GLS) technique. The difference between the two values is due to the GLS periodogram effectively using fewer data points (which correlates with less sensitivity) by down-weighting some of the runs with larger errors on the measured rate. Once the frequency dropped low enough, wherein the sample period exceeded the time span of the data, this detection threshold amplitude began to vary with frequency (as seen in the left-most plots of Figure~\ref{fig:Sensitivity}). In these regimes, only partial modulation waveforms are being captured in the time span of the data and the modulation phase began to have an affect on the results. If the phase was such that a mostly-linear portion of the modulation was captured (well between a peak and a trough in the underlying oscillation) then the sensitivity was reduced compared to other phase values.


%
%
\section{\label{sec:Summary}Summary and Conclusions}


The large, highly-pure sample of double-beta decays observed by the NEMO-3 collaboration provided a unique opportunity to probe the variability of a second-order, weak nuclear process. A power spectrum analysis was used to search for periodicities in the double-beta decay rate of \Mo based on $6.0195\,{\rm y}$ of data. Periodograms were generated using both the Lomb-Scargle technique (to be consistent with various previous searches for periodically varying decay rates), as well as its error-weighted extension, the Generalized Lomb-Scargle technique, with both resulting in power spectra consistent with the null hypothesis of no underlying modulation. 

This conclusion was reached by noting that the largest LS (GLS) power in the data-generated periodogram was exceeded 54.1\% (44.8\%) of the time by the largest LS (GLS) power from completely randomized, time-shuffled data sets. Furthermore, the frequencies at which these powers were found did not correlate with each other nor with any previously claimed periodicities at, or around, $1\,{\rm y^{-1}}$ for those relating to Earth's orbital period or those in the range $(10-15)\,{\rm y^{-1}}$ for those relating to solar synodic rotation rates~\cite{Jenkins:2008tt, Jenkins:2008vn, Fischbach:2009zz, Javorsek:2010sr, Sturrock:2010, Sturrock:2010bu, Sturrock:2014caa, Sturrock:2012gs, Jenkins:2012jc, Sturrock:2012re, Sturrock:2019dfx, parkhomov}. 
We estimate that the analyzed data set was sensitive to modulations with periods between one day and three years, if the relative amplitude of such modulations had exceeded approximately 2.0\% (2.5\%) when using the LS (GLS) technique. Although these constraints are an order of magnitude weaker than modulation searches in single-beta decay, nevertheless they represent the first ever limits for the second order process of double-beta decay.


\begin{acknowledgments}
We thank the staff of the Modane Underground Laboratory for their technical assistance in running the experiment. We acknowledge support by the grant agencies of the Czech Republic, CNRS/IN2P3 in France, RFBR in Russia (NCNIL No19-52-16002), APVV in Slovakia, the Science and Technology Facilities Council, part of U.K. Research and Innovation, and the NSF in the U.S.
\end{acknowledgments}



\end{document}